\documentclass{article}
\usepackage{spconf,amsmath,graphicx}

\title{Translation Operator in Graph Signal Processing: \\ A Generalized Approach}
\name{Amin Jalili, Sadid Sahami, and Chong-Yung Chi\thanks{This work was supported by the Ministry of Science and Technology, R.O.C., under Grant MOST 107-2221-E-007-021, MOST 108-2634-F-007-004 and MOST 108-2221-E-007-024.}}
\address{Institute of Communications Engineering, National Tsing Hua University, Hsinchu, Taiwan \\ \texttt{\small{amin.jalili@ieee.org, s.sahami@ieee.org, cychi@ee.nthu.edu.tw}}} 

\usepackage{graphicx}

\usepackage{multirow}

\usepackage{amsmath,amsthm,amssymb,amsthm,amsfonts}
\usepackage{dsfont}

\usepackage{fixltx2e}
\usepackage{ragged2e}
\usepackage{marvosym}
\usepackage{multido}
\usepackage{amssymb}
\usepackage{amsfonts}
\usepackage{cite}
\usepackage{booktabs}
\usepackage{multicol,lipsum}
\usepackage{mathtools}
\usepackage[outdir=./]{epstopdf}
\usepackage{mathtools}
\usepackage{eqnarray}
\usepackage{hyperref} 
\usepackage{textcomp} 
\usepackage{url}
\usepackage{booktabs,multirow,array}
\usepackage{tablefootnote}
\usepackage{enumerate}
\usepackage{afterpage}
\usepackage{bm}
\graphicspath{{Figs/}}

\newcommand{\mi}{\mathrm{i}} 
\usepackage{ragged2e}
\usepackage{stmaryrd}
\theoremstyle{plain}
\newtheorem{thm}{Theorem}

\newtheorem{prop}{Proposition}

\theoremstyle{definition}
\newtheorem{defn}{Definition}
\newtheorem{disc}{Discussion}

\newtheorem{remk}{Remark}

\makeatletter
\let\save@mathaccent\mathaccent
\newcommand*\if@single[3]{%
	\setbox0\hbox{${\mathaccent"0362{#1}}^H$}%
	\setbox2\hbox{${\mathaccent"0362{\kern0pt#1}}^H$}%
	\ifdim\ht0=\ht2 #3\else #2\fi
}
\newcommand*\rel@kern[1]{\kern#1\dimexpr\macc@kerna}
\newcommand*\widebar[1]{\@ifnextchar^{{\wide@bar{#1}{0}}}{\wide@bar{#1}{1}}}
\newcommand*\wide@bar[2]{\if@single{#1}{\wide@bar@{#1}{#2}{1}}{\wide@bar@{#1}{#2}{2}}}
\newcommand*\wide@bar@[3]{%
	\begingroup
	\def\mathaccent##1##2{%
		\let\mathaccent\save@mathaccent
		\if#32 \let\macc@nucleus\first@char \fi
		\setbox\z@\hbox{$\macc@style{\macc@nucleus}_{}$}%
		\setbox\tw@\hbox{$\macc@style{\macc@nucleus}{}_{}$}%
		\dimen@\wd\tw@
		\advance\dimen@-\wd\z@
		\divide\dimen@ 3
		\@tempdima\wd\tw@
		\advance\@tempdima-\scriptspace
		\divide\@tempdima 10
		\advance\dimen@-\@tempdima
		\ifdim\dimen@>\z@ \dimen@0pt\fi
		\rel@kern{0.6}\kern-\dimen@
		\if#31
		\overline{\rel@kern{-0.6}\kern\dimen@\macc@nucleus\rel@kern{0.4}\kern\dimen@}%
		\advance\dimen@0.4\dimexpr\macc@kerna
		\let\final@kern#2%
		\ifdim\dimen@<\z@ \let\final@kern1\fi
		\if\final@kern1 \kern-\dimen@\fi
		\else
		\overline{\rel@kern{-0.6}\kern\dimen@#1}%
		\fi
	}%
	\macc@depth\@ne
	\let\math@bgroup\@empty \let\math@egroup\macc@set@skewchar
	\mathsurround\z@ \frozen@everymath{\mathgroup\macc@group\relax}%
	\macc@set@skewchar\relax
	\let\mathaccentV\macc@nested@a
	\if#31
	\macc@nested@a\relax111{#1}%
	\else
	\def\gobble@till@marker##1\endmarker{}%
	\futurelet\first@char\gobble@till@marker#1\endmarker
	\ifcat\noexpand\first@char A\else
	\def\first@char{}%
	\fi
	\macc@nested@a\relax111{\first@char}%
	\fi
	\endgroup
}
\makeatother

\begin{document}
%
\maketitle
\begin{abstract}
	The notion of translation (shift) is straightforward in classical signal processing, however, it is challenging on an irregular graph structure. This paper aims to put forward an approach to characterize the abstract form of graph translation operator (GTO) by a natural generalization of abstract form of translation operators in classical domains. This approach yields to a very generic representation of GTO. Moreover, we show that the Schr\"{o}dinger equation, which describes the evolution of a dynamic system, intriguingly explains the idea behind translation on graph. Then we design an isometric translation operator in joint time-vertex domain consistent with the abstract form of translation operator in other signal domains.  
\end{abstract}

\section{Introduction}
\label{sec:intro}
Graph signal processing (GSP) generalizes the classical signal processing for analyzing \textit{structured data} in non-Euclidean spaces~\cite{Shuman2013Emerging,Sandryhaila2013Discrete,perraudin2017graph,Perraudin2017Stationary,Ortega2018Graph,Loukas2017JStationary,Shuman2016Vertex,Bronstein2017Geometric,perraudin2018Global,Giannakis2018Topology}. There are only a few research works that particularly addressed translation on  graph~\cite{Shuman2013Emerging,Sandryhaila2013Discrete,Girault2015Trans,Gavili2017Shift}. Shuman \textit{et al.}~\cite{Shuman2013Emerging,Shuman2016Vertex} defined the \textit{generalized translation operator} using the convolution of signal with the Kronecker delta function located at the target index. 
Sandryhaila and Moura~\cite{Sandryhaila2013Discrete} addressed the weighted adjacency matrix as the \textit{graph shift operator}. These operators lack isometry which is an essential property of a desired translation operator. Girault \textit{et al.}~\cite{Girault2015Trans} designed an isometric translation operator based on the graph Laplacian matrix. Moreover, Gavili \textit{et al.}~\cite{Gavili2017Shift} introduced graph shift operators based on the deformation of weighted adjacency matrix. Basically, operators can be considered as abstract mathematical objects with concrete manifestations in different domains. In other words, one may obtain an abstract characteristic of an operator in one domain and generalize it to other domains \textit{genetically inheriting a similar ``DNA"}. In this work, via an analytical approach, we shed some light on the characterization of generic form of graph translation operator (GTO) generalized from classical domains. Then, it will be shown that the resulting isometric GFT is closely related to the Schr\"{o}dinger equation expressing the evolution of a discrete dynamic system. Moreover, the joint time-vertex translation operator --- for short, joint translation operator (JTO) --- is proposed consistent with the characteristics of translation operators in continuous-time, discrete-time, and graph domains (cf.~Table~\ref{tab:tbl1}).

\textit{Notations:} Matrices and vectors are denoted by uppercase and lowercase boldface letters, $\bm{X}$ and $\bm{x}$, respectively. The $i$-th element of a vector is indexed by $\bm{x}[i]$, and the entry in $i$-th row and $j$-th column of a matrix is denoted by $\bm{X}[i,j]$. Then $\bm{X}^{\mathsf T}$, $\widebar{\bm{X}}$, and ${\bm{X}}^{*}$ stand for the transpose, conjugate, and adjoint (transposed complex conjugate) of the matrix $\bm{X}$, respectively. Moreover, ${\rm{vec}}(\bm{X})$ stands for the column vector by stacking all the columns of $\mathbf{X}$ sequentially and ${\rm{Diag}}(\bm{x})$ denotes a diagonal matrix by placing the elements of vector $\bm{x}$ on the main diagonal. Also, $\bf{I}$ and $\bf{1}$ denote the identity matrix and column vector of all ones, respectively. Symbols $\star$, $\otimes$, $\oplus$, $\odot$, and $\langle \cdot,\cdot\rangle$ represent the convolution operator, Kronecker product, Kronecker sum, Hadamard (element-wise) product, and inner product, respectively. Then $\mathbb{C}^{N\times N}$ and $\mathbb{R}^{N\times N}$ are the set of $N\times N$ complex and real matrices. Further $\mathbb{C}^{N}$ and $\mathbb{R}^{N}$ are the set of $N\times 1$ complex and real vectors. Finally, $\mi = \sqrt{-1}$ and $\llbracket a,b \rrbracket$ represents the integers between $a$ and $b$. 

\section{Background}
Let ${\mathsf G} \coloneqq (V, E, W)$ denote a fixed graph with finite vertex set $V$ with the cardinality $|V| = N$, $E = \{(i,j) | \; i,j \in V, j \sim i \} \subseteq V \times V$ is the edge set and $W: V \times V \rightarrow \mathbb{R}_{+}$ is a weight function. This function yields the weighted adjacency matrix as ${\mathbf W}_{\mathsf G} = [w_{i,j}] \in \mathbb{R}^{N \times N}$. Throughout this paper, we assume that the graph is weighted, connected and undirected. Then the graph Laplacian matrix is defined as $\mathbf{L}_{\mathsf G} \coloneqq {\rm{Diag}}\left({\mathbf W}{\mathbf 1}\right)~-~\mathbf W$.
A graph signal, represented by the vector $\bm{f} \in \mathbb{R}^{N}$, is defined as tying a scalar value to each node through the function $f_{\mathsf G}: V \rightarrow \mathbb{R}$ where $\bm{f}[i]$ is the function value at the vertex $i$. The graph Laplacian can be written as ${\mathbf{L}}_{\mathsf G} = \mathbf{\Psi}_{\mathsf G} {\mathbf{\Lambda}}_{\mathsf G} \mathbf{\Psi}_{\mathsf G}^{*}$ where $\mathbf{\Psi}_{\mathsf G}^{*}$ is the \textit{graph Fourier matrix} and ${\mathbf{\Lambda}}_{\mathsf G} = {\textrm{Diag}}\left(\left[\lambda_{0}, \ldots, \lambda_{N-1}\right]\right)$ is the eigenvalue matrix~\cite{Shuman2013Emerging}. The graph Fourier transform (GFT) and its inverse can be expressed as~\cite{Shuman2013Emerging}:  
$\widehat{\bm{x}}  = \mathcal{F}_{\mathsf G}\bm{x} = \mathbf{\Psi}_{\mathsf G}^{*} \bm{x}$ and $\bm{x} = \mathcal{F}_{\mathsf G}^{-1}\bm{x} = \mathbf{\Psi}_{\mathsf G} \widehat{\bm{x}}$, respectively,
where $\mathcal{F}_{\mathsf G}$ is the GFT operator and $\mathcal{F}_{\mathsf G}^{-1}$ accounts for the inverse GFT (IGFT) operator. 

\section{Translation Operator on Graph}
Let us begin with characterizing the abstract representation of translation operators in continuous-time domain as follows.  

\begin{remk}
	Let $x(t)$ be a continuous-time signal and $\hat{x}(f)$ be its Fourier transform. Let $\mathcal{T}_{\mathsf C}^{\tau}$ be the translation operator in continuous-time domain with translation value $\tau$ where $(\mathcal{T}_{\mathsf C}^{\tau}x)(t) = x(t-\tau)$. This can be formulated in the abstract form as (the proof is omitted here due to the limited space)
	\begin{align}\label{shiftOpContiDomain}
	\mathcal{T}_{\mathsf C}^{\tau} = \mathcal{F}_{\mathsf C}^{-1}\mathcal{M}_{\mathsf C}^{\mathsf \tau}\mathcal{F}_{\mathsf C},
	\end{align}
	where $\mathcal{M}_{\mathsf C}^{\tau} = \exp{(-\mi 2\pi f \tau)}$ and $\mathcal{F}_{\mathsf C}$ is the Fourier transform operator in continuous-time domain. Moreover, one can write $(\mathcal{M}\hat{x})(f) \coloneqq 2\pi f \hat{x}(f)$. 
\end{remk}

Now, we take a quick look at the translation operator in discrete-time domain. The discrete Fourier transform (DFT) operator $\mathcal{F}_{\mathsf D}$ and its inverse $\mathcal{F}_{\mathsf D}^{-1}$ can be represented in a matrix form as~\cite{vetterli2014Foundations}: $\widehat{\mathbf{x}} = \mathcal{F}_{\mathsf D}{\mathbf{x}} = \mathbf{\Psi}_{\mathsf D}^{*} {\mathbf{x}}$ and $
{\mathbf{x}} = \mathcal{F}_{\mathsf D}^{-1}{\widehat{\mathbf{x}}} = {\mathbf{\Psi}_{\mathsf D}}\widehat{\mathbf{x}}
$, respectively, where ${\mathbf x}$ is the signal vector,
$\mathbf{\Psi}_{\mathsf D}[n,k] =  e^{\mi \omega_{k}n}$, and $\omega_{k} \coloneqq 2\pi(k-1)/M$ for all $n,k \in {\llbracket 1,M \rrbracket}$.

\begin{defn}
	Let $x[n], \; n\in {\llbracket 1,M \rrbracket}$ be a discrete-time signal and the right-circular translation operator $\mathcal{T}_{\mathsf D}^{\upsilon}$ in discrete-time domain with the translation value $\upsilon$ is defined as $(\mathcal{T}_{\mathsf D}^{\upsilon}x)[n] \coloneqq x[n-\upsilon]$. Let ${\mathbf x}\coloneqq \left[x[1], x[2], \ldots, x[M]\right]^{\mathsf T}$ be the signal in vector form. The unit translation in discrete-time domain, simply denoted by $\mathcal{T}_{\mathsf D}$, can be expressed as $\mathcal{T}_{\mathsf D}\mathbf{x} = \mathbf{T}_{\mathsf D}\mathbf{x} =  [\mathbf{e}_2, \mathbf{e}_3, \ldots, \mathbf{e}_{M}, \mathbf{e}_{1}] \mathbf{x}$
	where $\mathbf{e}_i$ is the $M\times 1$ unit vector with the $i$-th entry equal to 1. 
\end{defn}

By a classic interpretation, discrete-time domain can be modeled as the $M$-Cycle graph $\mathsf D$ with all unit edge weights. Clearly, one can write the weighted adjacency matrix of graph $\mathsf D$ as ${\mathbf W}_{\mathsf D} = \mathbf{T}_{\mathsf D}$. 

\begin{remk}
	The translation operator in the discrete-time domain can be expressed as
	\begin{align}\label{DiscreteDiffOp}
		\mathcal{T}_{\mathsf D}^{\upsilon} = \mathcal{F}_{\mathsf D}^{-1} \mathcal{M}_{\mathsf D}^{\upsilon} \mathcal{F}_{\mathsf D},
	\end{align}	
	where $\mathcal{F}_{\mathsf D}$ is the DFT operator, $\mathcal{M}_{D}^{\upsilon} = e^{-\mi\upsilon\mathbf{M}_{\mathsf D}}$ and $\mathbf{M}_{\mathsf D} \coloneqq {\rm{Diag}}\big([\omega_{0}, \ldots, \omega_{M-1}]\big)$ is the diagonal matrix containing angular frequencies in the discrete-time domain as $\omega_{k} \coloneqq 2\pi (k-1)/M$ for all $k \in \llbracket 0,M-1 \rrbracket$. The matrix representation of $\upsilon$-translation can be written as 
	\begin{align}\label{DTtransMat}
		\mathbf{T}_{\mathsf D}^{\upsilon} = {\mathbf{\Psi}_{\mathsf D}} \exp(-\mi \upsilon\mathbf{M}_{\mathsf D}) \mathbf{\Psi}_{\mathsf D}^{*}.
	\end{align}
\end{remk} 

In the following, we will discuss the generic representation of translation operator on graph. Girault \textit{et. al.}~\cite{Girault2015Trans,girault2015Signal} are the first who introduced the isometric GTO. To be specific, they designed their operator based on the properties of isometry and convolutivity which led to the following general form $\mathbf{T}_{\mathsf G} = \exp{(\mi \mathbf{\Omega})}$ for which the matrix $\mathbf{\Omega}$ has to be specified (cf.~\cite[Eq.~6]{Girault2015Trans}). However, since the translation operators in continuous-time and discrete-time domains (cf.~\eqref{shiftOpContiDomain} and~\eqref{DiscreteDiffOp}) are both \textit{isometric} and \textit{convolutive}, it is indeed not necessary for such design. In other words, one may characterize the abstract form of GTO $\mathcal{T}_{\mathsf G}$ simply by generalizing from the classical domains as follows
\begin{align}\label{shiftGraph}
\mathcal{T}_{\mathsf G}^{\kappa} \coloneqq \mathcal{F}_{\mathsf G}^{-1} \mathcal{M}_{\mathsf G}^{\kappa} \mathcal{F}_{\mathsf G}, 
\end{align}
where $\kappa$ is the translation value, $\mathcal{M}_{\mathsf G}^{\kappa} = \exp{\left(-\mi \kappa {\mathbf{M}}_{\mathsf G}\right)}$ and ${\mathbf{M}}_{\mathsf G}$ is a diagonal matrix containing the angular frequencies in the graph setting and $\mathcal{F}_{\mathsf G}$ accounts for the GFT operator. 

\begin{figure*}[t!]
	\hspace*{4mm}
	\includegraphics[scale=.62]{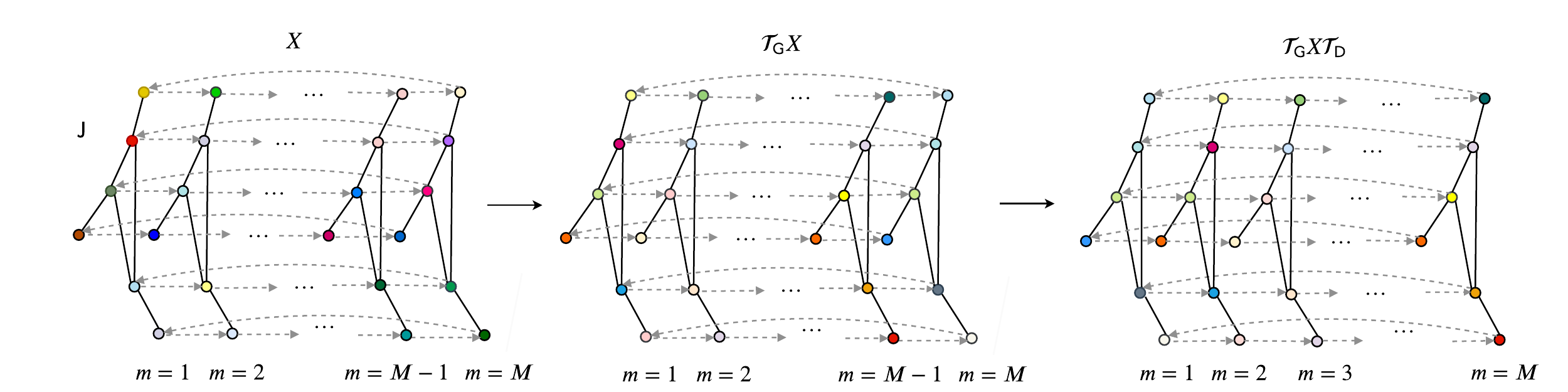}
	\caption{Illustration of the joint time-vertex translation where colored circles denote the signal values on joint graph $\mathsf J$.}
	\label{fig:JTOfig}
\end{figure*}

\begin{disc}\label{disGGTO}
	\normalfont
	The generic representation~\eqref{shiftGraph} allows one either to use weighted adjacency matrix $\mathbf{W}$ or graph Laplacian $\mathbf{L}_{\mathsf G}$ for defining the GTO. As the matrix $\mathbf{M}_{\mathsf G}$ is assigned, the operator $\mathcal{T}_{\mathsf G}$ is then well-defined. Then, we discuss some examples as the manifestations for abstract form of GTO. \\
	(i) The notion of graph frequency is defined in an analogous manner to the frequency in the continuous domain such that 
	$
	-\Delta e^{\mi 2\pi ft} = (2\pi f)^2 e^{\mi2\pi ft}
	$
	where $\Delta$ is the Laplace-Beltrami operator and $e^{\mi2\pi ft}$ for $f\in \mathbb{R}$ are its eigenfunctions. Moreover, the (combinatorial) graph Laplacian can be considered as an approximation of the Laplace-Beltrami operator up to a negative sign (i.e., $-\Delta$)~\cite{HAMMOND2011129Wavelets}. Following these observations from the continuous space, Shuman \textit{et al.}~\cite{Shuman2013Emerging} specified that $\lambda_{\ell}$ for $\ell \in {\llbracket 0,N-1 \rrbracket}$ carries the \textit{frequency notion} in graph setting. Then the equivalent of angular frequencies in graph setting can be defined as $\{\varpi_{\ell} \coloneqq \sqrt{\lambda_{\ell}}, \ell \in {\llbracket 0,N-1 \rrbracket}\}$ as a natural generalization from the continuous space to graph setting. Then we define
	$
	\mathbf{M}_{\mathsf G} \coloneqq {\rm{Diag}}\left(\left[\varpi_{0},, \ldots, \varpi_{N}\right]\right)
	$
	and the matrix representation of generalized GTO with $\kappa$-translation can be written as 
	\begin{align}\label{GGTO}
	\mathbf{T}_{\mathsf G}^{\kappa} = \mathbf{\Phi}_{\mathsf G} \exp(-\mi \kappa \mathbf{M}_{\mathsf G}) \mathbf{\Phi}_{\mathsf G}^{*}.
	\end{align}
	(ii) Considering $\mathcal{F}_{\mathsf G} = \mathbf{\Phi}_{\mathsf G}$ and 
	$
	\mathbf{M}_{\mathsf G} \coloneqq {\rm{Diag}}\left(\left[\widetilde{w}_{0}, \ldots, \widetilde{w}_{N-1}\right]\right)
	$,
	where $\widetilde{w}_{\ell} = \pi \sqrt{\lambda_{\ell}/\rho}$ for $\ell \in {\llbracket 0,N-1 \rrbracket}$ and $\rho$ is upper bound on the eigenvalues of graph Laplacian matrix, gives rise to the GTO defined by Girault \textit{et al.}~\cite{Girault2015Trans} where they proposed the notion of \textit{reduced graph frequencies} $\widetilde{w}_{\ell}$ such that the eigenvalues are mapped to the interval $[0,\pi]$. \\
	(iii) Using the eigenvector basis of weighted adjacency matrix $\mathbf{W}$ as the matrix representation of $\mathcal{F}_{\mathsf G}$ and defining  
	$
	\mathbf{M}_{\mathsf G} \coloneqq {\rm{Diag}}\big([0, 2\pi/N, \ldots, 2\pi(N-1)/N]\big)
	$
	,~\eqref{shiftGraph} results in the Gavili \textit{et al.}'s graph shift operator (cf.~\cite{Gavili2017Shift}, denoted by $\mathbf{A}_e$). \\
	(iv) Considering the eigenvector basis of $\mathbf{W}$ as the matrix representation of $\mathcal{F}_{\mathsf G}$ and defining    
	$
	\mathbf{M}_{\mathsf G} \coloneqq {\rm{Diag}}\big([\phi_0, \ldots, \phi_{N-1}]\big)
	$
	where $\phi_\ell, \forall \ell \in {\llbracket 0,N-1 \rrbracket}$ as the arbitrary phases in $[0,2\pi]$ and $\phi_\ell \neq \phi_k$ for $\ell \neq k$,~\eqref{shiftGraph} leads to the Gavili \textit{et al.}'s graph shift operator (cf.~\cite{Gavili2017Shift}, denoted by $\mathbf{A}_\phi$). \hfill $\Box$
\end{disc}

After the discussion about the characterization of the translation operator in various domains, in the next remark, we describe the intuition behind translation graph.

\begin{remk}
	Consider a dynamic $N$-state system defined on the connected graph $\mathsf G$ where the state in \textit{evolution-time} $t \in \mathbb{R}_{+}$ is described by a column vector ${\bm{u}}(t)$. The \textit{Schr\"{o}dinger equation} is expressed as
	\begin{equation}\label{SchrodEq}
	(\mi \alpha\partial_t - {\boldsymbol{\mathsf{H}}}_{\mathsf G}) {\bm u}(t) = 0,
	\end{equation}
	where $\partial_t$ is the partial derivative with respect to the evolution-time, $\alpha$ is a constant --- \textit{in the original equation, it is the Plank's reduced constant}, and ${\bm{u}}(0)$ is the initial state. In the context of GSP, ${\bm{u}}(0) \in \mathbb{R}^{N}$ corresponds to the given graph signal. Here, ${\boldsymbol{\mathsf{H}}}_{\mathsf G}$ is any self-adjoint matrix representing the characteristic of graph ${\mathsf G}$ called \textit{Hamiltonian}. Suppose ${\boldsymbol{\mathsf{H}}}_{\mathsf G} = \mathbf{\Psi}_{\mathsf G}\mathbf{M}_{\mathsf G}\mathbf{\Psi}_{\mathsf G}^*$. Then one can obtain the solution of~\eqref{SchrodEq} as
	\begin{align*}
	\bm{u}(t) = e^{-\mi t {\boldsymbol{\mathsf{H}}}_{\mathsf G}/\alpha} \bm{u}(0) = \sum_{k=0}^{N-1} e^{-\mi t \gamma_{k}/\alpha} \langle \bm{u}(0),{\bm  \psi}_{{\mathsf G}, k} \rangle {\bm \psi}_{{\mathsf G}, k},
	\end{align*}
	where ${\bm \psi}_{{\mathsf G}, k}$ is the $k$-th column of $\mathbf{\Psi}_{\mathsf G}$ and $\gamma_{k}$ is the $k$-th entry on the main diagonal of $\mathbf{M}_{\mathsf G}$ (corresponding to some angular frequency). This identity represents the evolution of graph signal $\bm{u}(0)$ on graph $\mathsf G$ on the continuous evolution time axis. Then the \textit{transition function}~\cite{Coutinho2014Quantum} on graph $\mathsf G$ is defined as 
	\begin{align}\label{transFn}
	H_{\mathsf G}(t) \coloneqq \exp\left(-\mi t {\boldsymbol{\mathsf{H}}}_{\mathsf G}/\alpha\right)
	= \sum_{r=0}^{\infty} \frac{(-\mi t/\alpha)^r}{r!} {\boldsymbol{\mathsf{H}}}_{\mathsf G}^{r}, 
	\end{align}
	which is a matrix function presenting the evolution of continuous time quantum walk over $\mathsf G$. It is interesting to observe that, \textit{for integer values of $t$}, the isometric GTO is equivalent to the transition function (where the translated graph signal is equivalent to the evolutionized form of the graph signal).  
\end{remk} 

\section{Joint Translation Operator}
A time-varying graph signal, represented by the matrix 
$
{\bm{X}} = [{\bm{x}}_{1}, {\bm{x}}_{2}, \ldots,{\bm{x}}_{M}] \in \mathbb{R}^{N \times M}
$
where ${\bm{x}}_{m}$ denotes the graph signal at time $m \in {\llbracket 1,M \rrbracket}$ (as illustrated in the leftmost part in Fig.~\ref{fig:JTOfig}). The joint Fourier transform (JFT) is defined as
$
\widehat{{\bm{X}}} \coloneqq \mathbf{\Psi}_{\mathsf G}^{*} {\bm{X}} {\widebar{\mathbf{\Psi}}_{\mathsf D}}
$
where $\mathbf{\Psi}_{\mathsf G}$ and $\mathbf{\Psi}_{\mathsf D}$ are the GFT and DFT matrices, respectively~\cite{Loukas2016Frequency}. The JFT coefficient of ${\bm{X}}$ corresponding to the joint angular frequency $(\varpi_{\ell},\omega_{k})$ is denoted by $\widehat{{\bm{X}}}[\ell,k]$ where $\varpi_{\ell} = \sqrt{\lambda_{\ell}}$ and $\omega_{k} = 2\pi k/M$ are the $\ell$-th and $k$-th angular frequencies in graph setting and discrete-time domain, respectively. JFT and its inverse can be rewritten as~\cite{Loukas2016Frequency} 
$
\widehat{\bm{x}} = \mathcal{F}_{\mathsf J}\bm{x} =\mathbf{\Psi}_{\mathsf J}^{*} \bm{x}
$
and
$
\bm{x} = \mathcal{F}_{\mathsf J}^{-1}\widehat{\bm{x}} = \mathbf{\Psi}_{\mathsf J} \widehat{\bm{x}}
$
, respectively, where ${\bm{x}} = {\rm{vec}}({\bm{X})}$ and ${\mathbf{\Psi}_{\mathsf J}} \coloneqq {\mathbf{\Psi}_{\mathsf D}} \otimes {\mathbf{\Psi}_{\mathsf G}}$ is a unitary matrix. Besides, using the notion of joint filtering~\cite{Grassi2018Time}, an operator in joint time-vertex domain is convolutive if it can be written as ${\mathbf{H}_\mathsf{J}} = \mathbf{\Psi}_{\mathsf J} {\widehat{\mathbf{H}}_\mathsf{J}} \mathbf{\Psi}_{\mathsf J}^{*}$ where ${\widehat{\mathbf{H}}_\mathsf{J}}$ is a diagonal matrix.

\begin{table*}[t!]
	\caption{Characteristics of translation operators in various signal domains} 
	\label{tab:tbl1}
	\centering 
		\begin{tabular}{@{}l l l@{}} 
			\toprule[1pt]
			\textbf{Domain} & \textbf{Abstract form} & \textbf{Description}   \\ 
			\hline
			\multirow{2}{*}{Continuous-time (cf.~\eqref{shiftOpContiDomain})}  & \multirow{2}{*}{$
			\mathcal{T}_{\mathsf C}^{\tau} = \mathcal{F}_{\mathsf C}^{-1}\mathcal{M}_{\mathsf C}^{\mathsf \tau}\mathcal{F}_{\mathsf C}
			$}	
			& $\mathcal{F}_{\mathsf C}$: Continuous-time Fourier transform operator, $\mathcal{M}_{\mathsf C}^{\tau} = \exp{(-\mi 2\pi f\tau)}$,\\ 
			& & Angular frequency multiplication operator:  $(\mathcal{M}\hat{x})(f) \coloneqq (2\pi f\hat{x})(f)$
			\\
			\hline
			\multirow{2}{*}{Discrete-time (cf.~\eqref{DiscreteDiffOp})}  & \multirow{2}{*}{$
			\mathcal{T}_{\mathsf D}^{\upsilon} = \mathcal{F}_{\mathsf D}^{-1} \mathcal{M}_{\mathsf D}^{\upsilon} \mathcal{F}_{\mathsf D}
			$}	
			& $\mathcal{F}_{\mathsf D}$: DFT operator, $\mathcal{M}_{D}^{\upsilon} \coloneqq \exp\left({-\mi\upsilon\mathbf{M}_{\mathsf D}}\right)$, \\ 
			& & $\mathbf{M}_{\mathsf D}$: Diagonal matrix of discrete angular frequencies  
			\\
			\hline
			\multirow{2}{*}{Graph setting (cf.~\eqref{shiftGraph})}   & \multirow{2}{*}{$
			\mathcal{T}_{\mathsf G}^{\upsilon} \coloneqq \mathcal{F}_{\mathsf G}^{-1} \mathcal{M}_{\mathsf G}^{\kappa} \mathcal{F}_{\mathsf G}
			$}	
			& $\mathcal{F}_{\mathsf G}$: GFT operator, $\mathcal{M}_{\mathsf G}^\kappa \coloneqq \exp{\left(-\mi \kappa {\mathbf{M}}_{\mathsf G}\right)}$, \\ 
			& & ${\mathbf{M}}_{\mathsf G}$: Diagonal matrix of angular frequencies in graph setting 
			\\
			\hline
			\multirow{3}{*}{Joint time-vertex (cf.~\eqref{JTVtransAbstract})}  & \multirow{3}{*}{$
			\mathcal{T}_{\mathsf J}^{(\kappa,\upsilon)} = \mathcal{F}_{\mathsf J}^{-1} \mathcal{M}_{\mathsf J}^{(\kappa,\upsilon)} \mathcal{F}_{\mathsf J}
			$}	
			& $\mathcal{F}_{\mathsf J}$: JFT operator, $\mathcal{M}_{\mathsf J}^{(\kappa,\upsilon)} \coloneqq \exp\left(-\mi \mathbf{M}_{\mathsf J}^{(\kappa,\upsilon)}\right)$, \\ 
			& & $\mathbf{M}_{\mathsf J}^{(\kappa,\upsilon)}$: Diagonal matrix of joint angular frequencies 
			\\
			\bottomrule[1pt]
		\end{tabular}
\end{table*}


\begin{defn}\label{JTVtranOper}
	We define the translation of joint time-vertex signal by applying the graph translation $\mathbf{T}_{\mathsf G}^{\kappa}$ on the graph dimension and the discrete-time translation\footnote{Here, $\mathbf{T}_{\mathsf D}$ accounts for the right-circular translation operator (with unit shift value) in the discrete-time domain. If we consider $\mathbf{x} = [x_{0}, x_{1}, \ldots, x_{N-1}]^{\mathsf T}$ as the discrete-time signal, then $\mathbf{x}^{\mathsf T} \mathbf{T}_{\mathsf D}^{\mathsf T} = (\mathbf{T}_{\mathsf D} \mathbf{x})^{\mathsf T} = [x_{N-1}, x_{0}, x_{1}, \ldots, x_{N-2}]^{\mathsf T}$ is actually the right-circular translation of $\mathbf{x}^{\mathsf T}$.} $(\mathbf{T}_{D}^{\mathsf T})^{\upsilon}$ along the time axis as  
	$
	{\bm{X}}^{(\kappa,\upsilon)} \coloneqq \mathbf{T}_{\mathsf G}^{\kappa} {\bm{X}} \left(\mathbf{T}_{\mathsf D}^{\mathsf T}\right)^{\upsilon}
	$
	where $\kappa$ and $\upsilon$ account for the translation values in the graph setting and discrete-time domain, respectively\footnote{Without loss of generality, we use the generic form of GTO (cf.~\eqref{GGTO}). However, any manifestation of the abstract representation~\eqref{GGTO} can be exploited to define the JTO.}. Then the joint time-vertex translation operator $\mathcal{T}_{\mathsf J}^{(\kappa,\upsilon)}$ can be defined as 
	$
	{\bm{x}}^{(\kappa,\upsilon)} = \mathcal{T}_{\mathsf J}^{(\kappa,\upsilon)}{\bm{x}}
	$
	where ${\bm{x}}^{(\kappa,\upsilon)} = {\rm{vec}}\left({\bm{X}}^{(\kappa,\upsilon)}\right)$, ${\bm{x}} = {\rm{vec}}({\bm{X}})$ and the matrix representation of $\mathcal{T}_{\mathsf J}^{(\kappa,\upsilon)}$ can be obtained as\footnote{This is obtained using the following property of Kronecker product: for any given matrices $\mathbf{A} \in \mathbb{F}^{m \times m}$, $\mathbf{B} \in \mathbb{F}^{n \times n}$, and $\mathbf{X} \in \mathbb{F}^{m \times n}$, where $\mathbb{F}$ is any field, the equation $\mathbf{B} {\mathbf{X}} \mathbf{A} = \mathbf{S}$ can be written as $(\mathbf{A}^{\mathsf T} \otimes \mathbf{B}){\rm{vec}}({\mathbf{X}}) = {\rm{vec}}(\mathbf{S})$~\cite[Proposition~12.1.4]{Lancaster1985theory}.}.
	\begin{align}\label{JTVtransMat}
	\mathbf{T}_{\mathsf J}^{(\kappa,\upsilon)} = \mathbf{T}_{\mathsf D}^{\upsilon} \otimes \mathbf{T}_{\mathsf G}^{\kappa}.
	\end{align}
\end{defn}

In a special framework, the joint time-vertex domain is modeled as the multilayer graph $\mathsf J$ (namely, \textit{joint graph}) resulting from the Cartesian product of $\mathsf G$ and $\mathsf D$~\cite{Grassi2018Time}. We use $\mathsf J$ to present the idea behind our definition of translation in joint time-vertex domain in Fig.~\ref{fig:JTOfig} where $\kappa = \upsilon = 1$.

\begin{prop}
	\normalfont
	The joint time-vertex translation operator $\mathcal{T}_{\mathsf J}^{(\kappa,\upsilon)}$ is a unitary operator.
	
	\textit{Proof:} 
	\normalfont{
		It is sufficient to prove it for the unit joint time-vertex translation simply denoted by $\mathbf{T}_{\mathsf J}$. Then we have
		\begin{align}
		\begin{split}
		\mathbf{T}_{\mathsf J}\mathbf{T}_{\mathsf J}^{*} 
		&= \left(\mathbf{T}_{\mathsf D} \otimes \mathbf{T}_{\mathsf G}\right) \left(\mathbf{T}_{\mathsf D} \otimes \mathbf{T}_{\mathsf G}\right)^{*} \\
		&= \left(\mathbf{T}_{\mathsf D} \mathbf{T}_{\mathsf D}^{*}\right) \otimes \left(\mathbf{T}_{\mathsf G} \mathbf{T}_{\mathsf G}^{*}\right)
		= \mathbf{I}_{M} \otimes \mathbf{I}_{N} = \mathbf{I}_{NM},
		\end{split}
		\end{align}
		where the third equality holds since $\mathbf{T}_{\mathsf G}$ and $\mathbf{T}_{\mathsf D}$ are unitary matrices. Similarly, one can show that $\mathbf{T}_{\mathsf J}^{*}\mathbf{T}_{\mathsf J} = \mathbf{I}_{NM}$. \hfill $\blacksquare$}
\end{prop}

\begin{thm}\label{theoremJTVtrans}
	The proposed joint time-vertex translation operator $\mathcal{T}_{\mathsf J}^{(\kappa,\upsilon)}$ can be written in the same abstract representation of translation operators in the continuous-time (cf.~\eqref{shiftOpContiDomain}) and discrete-time (cf.~\eqref{DiscreteDiffOp}) domains as follows 
	\begin{align}\label{JTVtransAbstract}
	\mathcal{T}_{\mathsf J}^{(\kappa,\upsilon)} = \mathcal{F}_{\mathsf J}^{-1} \mathcal{M}_{\mathsf J}^{(\kappa,\upsilon)} \mathcal{F}_{\mathsf J},
	\end{align}
	where 
	$
	\mathcal{M}_{\mathsf J}^{(\kappa,\upsilon)} \coloneqq \exp\left(-\mi \mathbf{M}_{\mathsf J}^{(\kappa,\upsilon)}\right)
	$
	such that
	$
	\mathbf{M}_{\mathsf J}^{(\kappa,\upsilon)} \coloneqq 
	\upsilon \mathbf{M}_{\mathsf D}\oplus \kappa \mathbf{M}_{\mathsf G} =
	\begin{bmatrix}
	\zeta_{i,j}^{(\kappa,\upsilon)}
	\end{bmatrix}$ 
	and $\zeta_{i,j}^{(\kappa,\upsilon)} \coloneqq \kappa \varpi_{i} + \upsilon \omega_{j}$ for all $i~\in~{\llbracket 0,N-1 \rrbracket}$, $j\in {\llbracket 0,M-1 \rrbracket}
	$. \\
	
	Proof: 
	\normalfont{Following from~\eqref{JTVtransMat}, one can write
		\begin{align*}\label{JTVtranslation}
		\begin{split}
		\mathbf{T}_{\mathsf J}^{(\kappa,\upsilon)} 
		&= \mathbf{T}_{\mathsf D}^{\upsilon} \otimes \mathbf{T}_{\mathsf G}^{\kappa} 
		= \left(\mathbf{\Psi}_{\mathsf D} e^{-\mi \upsilon\mathbf{M}_{\mathsf D}} \mathbf{\Psi}_{\mathsf D}^{*}\right) \otimes \left(\mathbf{\Psi}_{\mathsf G} e^{-\mi \kappa \mathbf{M}_{\mathsf G}} \mathbf{\Psi}_{\mathsf G}^{*}\right) \\
		& = \left(\mathbf{\Psi}_{\mathsf D} \otimes \mathbf{\Psi}_{\mathsf G}\right) \left(e^{-\mi \upsilon \mathbf{M}_{\mathsf D}} \otimes e^{-\mi \kappa \mathbf{M}_{\mathsf G}} \right) \left(\mathbf{\Psi}_{\mathsf D} \otimes \mathbf{\Psi}_{\mathsf G}\right)^{*} \\
		&= \mathbf{\Psi}_{\mathsf J}  
		e^{-\mi(\upsilon \mathbf{M}_{\mathsf D}\oplus \kappa \mathbf{M}_{\mathsf G})} \mathbf{\Psi}_{\mathsf J}^{*}
		= \mathbf{\Psi}_{\mathsf J} \exp\left({-\mi \mathbf{M}_{\mathsf J}^{(\kappa,\upsilon)}}\right) \mathbf{\Psi}_{\mathsf J}^{*}.
		\end{split}
		\end{align*} 
		Then this can be written in an abstract form as~\eqref{JTVtransAbstract} considering $\mathbf{\Psi}_{\mathsf J}^{*}$ as the matrix representation of $\mathcal{F}_{\mathsf J}$ and $\mathcal{M}_{\mathsf J}^{(\kappa,\upsilon)}$ as defined in Theorem~\ref{theoremJTVtrans}.  \hfill $\blacksquare$
	}
\end{thm}

\begin{disc}
	\normalfont
	There are two critical points behind the JTO as follows. First, by considering different matrix representations of the GTO as $\mathcal{T}_{\mathsf G}^{(\kappa,\upsilon)}$ (defined by $\mathcal{F}_{\mathsf G}$ and $\mathcal{M}_{\mathsf G}$ --- cf.~\eqref{shiftGraph} and Discussion~\autoref{disGGTO}), the abstract form~\eqref{JTVtransAbstract} leads to different manifestations of JTO as $\mathcal{T}_{\mathsf J}^{(\kappa,\upsilon)}$ in the matrix form. This shows the very generic nature of the proposed isometric JTO (cf.~~\eqref{JTVtransAbstract}). Second, by assuming $\kappa = \upsilon$, $\mathcal{T}_{\mathsf J}^{(\kappa,\upsilon)}$ reduces to a special case as the GTO on joint graph $\mathsf J$ (this can be defined based on the joint Laplacian matrix corrsponding to $\mathsf J$ using~\eqref{GGTO}). Therefore, our proposed JTO is \textit{more general than defining GTO on the joint graph $\mathsf J$}\footnote{This plays a key role for defining stationarity in joint time-vertex domain via JTO. Indeed, it yields a \textit{more general notion of stationarity} in joint time-vertex domain than stationarity on joint graph $\mathsf J$ as follows (this matter is also elaborated by~\cite{Loukas2017JStationary}, where they defined joint stationarity based on the notion of joint filtering). A joint time-vertex stochastic signal $\bm{x}$ on graph $\mathsf{G}$ is called joint time-vertex wide sense stationary if and only if \textit{for all $\kappa$ and $\upsilon$}:\\
		(1) $\mathbb{E}\left[\left(\mathcal{T}_{\mathsf J}^{(\kappa,\upsilon)} \bm{x}\right)\right] = \mathbb{E}[\bm{x}]$; \\
		(2) $\mathbb{E}\left[\left(\mathcal{T}_{\mathsf J}^{(\kappa,\upsilon)} \bm{x}\right)\left(\mathcal{T}_{\mathsf J}^{(\kappa,\upsilon)} \bm{x}\right)^{*}\right] = \mathbb{E}[\bm{x}\bm{x}^{*}]$, \\
		where $\mathbb{E}[\cdot]$ accounts for the statistical expectation.}. 
	\hfill $\Box$
\end{disc}

\begin{prop}
	\normalfont
	The properties of JTO, given by~\eqref{JTVtransAbstract}, are: 
	\vspace*{-0.15cm}
	\begin{enumerate}[(i)]
		\setlength{\itemsep}{-4pt}
		\item It is linear, convolutive (since $\mathbf{M}_{\mathsf J}^{(\kappa,\upsilon)}$ is a diagonal matrix) and isometric (because it is a unitary operator). 
		\item $\mathcal{T}_{\mathsf J}^{(\kappa,\upsilon)}$s commute with each other as
		$
		\mathcal{T}_{\mathsf J}^{(\kappa_{1},\upsilon_{1})}\mathcal{T}_{\mathsf J}^{(\kappa_{2},\upsilon_{2})} = \mathcal{T}_{\mathsf J}^{(\kappa_{2},\upsilon_{2})} \mathcal{T}_{\mathsf J}^{(\kappa_{1},\upsilon_{1})} = \mathcal{T}_{\mathsf J}^{(\kappa_{1}+\kappa_{2},\upsilon_{1}+\upsilon_{2})}.
		$
		\item The power spectrum of time varying graph signal signal ${\bm{X}}$ is invariant under the operator $\mathcal{T}_{\mathsf J}^{(\kappa,\upsilon)} $ as $\Big|{\widehat{\bm{X}}}^{(\kappa,\upsilon)}[\ell,k]\Big|^2 = \Big|{\widehat{\bm{X}}}[\ell,k]\Big|^2$ for all $\ell \in {\llbracket 1,N \rrbracket}$ and $k\in {\llbracket 1,M \rrbracket}$.
		\item The set of $\mathcal{Z} \coloneqq \{\mathcal{T}_{\mathsf J}^{(\kappa,\upsilon)}: \kappa,\upsilon \in \mathbb{Z}_{+}\}$, with the operation of multiplication, forms a mathematical translation abelian group ($\mathbb{Z}_{+}$ is the set of nonnegative integers).
	\end{enumerate}
\end{prop}

We also discuss on the joint shift operator defined by~\cite{SEGARRA2018Statistical} as follows.

\begin{disc}
	Segarra \textit{et al.}~\cite{SEGARRA2018Statistical} defined shift operator in joint time domain as follows
	\begin{align*}
		\mathcal{S}_{\mathsf J} \coloneqq \mathbf{W}_{\mathsf D}\oplus \mathbf{W}_{\mathsf G} = \left(\mathbf{\Phi}_{\mathsf D} \otimes \mathbf{\Phi}_{\mathsf G}\right) 
		(\mathbf{\Gamma}_{\mathsf D} \oplus \mathbf{\Gamma}_{\mathsf G})
		\left(\mathbf{\Phi}_{\mathsf D} \otimes \mathbf{\Phi}_{\mathsf G}\right)^{*}
	\end{align*}
	where $\mathbf{W}_{\mathsf D} = \mathbf{\Phi}_{\mathsf D}\mathbf{\Gamma}_{\mathsf D} \mathbf{\Phi}_{\mathsf D}$ and $\mathbf{W}_{\mathsf G} = \mathbf{\Phi}_{\mathsf G} \mathbf{\Gamma}_{\mathsf G}\mathbf{\Phi}_{\mathsf G}^{*}$ are the weighted adjacency matrices, as the  shift operators, in discrete-time and graph domains, respectively. There are crucial points behind this operator as follows: (1) In general, this is \textit{not isometric} which is a crucial property for a shift operator; (2) This is a \textit{univariate operator} and indeed is the weighted adjacency matrix of joint graph $\mathsf J$. This definition treats discrete-time and graph domains \textit{equally}, however, it does not include the feasible shifts with different values in the two domains. This is of significant importance for defining stationarity such that the defined stationarity based on this shift operator is a special case of stationarity in joint time-vertex domain. We suggest the following reformulation for this operator to be a bivariate operator as follows
	\begin{align}
		\mathcal{S}_{\mathsf J}^{(\kappa,\upsilon)} \coloneqq  \left(\mathbf{\Phi}_{\mathsf D} \otimes \mathbf{\Phi}_{\mathsf G}\right) 
		(\mathbf{\Gamma}_{\mathsf D}^{\upsilon} \oplus \mathbf{\Gamma}_{\mathsf G}^{\kappa})
		\left(\mathbf{\Phi}_{\mathsf D} \otimes \mathbf{\Phi}_{\mathsf G}\right)^{*}.
	\end{align}
	Then one can define the joint stationarity via this operator. 
	\hfill $\Box$
\end{disc}
Finally, Table~\ref{tab:tbl1} summarizes the abstract representations of isometric translation operators in various signal domains where they share similar structural characteristics. 

\section{Conclusion}
We have presented an approach for characterizing the generic form of translation operator on graph. Different matrix representations of the abstract form lead to different manifestations of GTO including all the existing isometric GTOs as special cases. Moreover, we showed the connection between translation on graph and the time evolution of a dynamic system modeled by the Schr\"{o}dinger equation. Then we designed the translation operator in joint time-vertex domain in harmony with the abstract form of translation operators in other signal domains (cf.~Table~\ref{tab:tbl1}). The proposed isometric JTO paves the way for studying the stationarity in time-vertex domain via translation invariance, which is our on-going research work.


\let\oldthebibliography=\thebibliography
\let\endoldthebibliography=\endthebibliography
\renewenvironment{thebibliography}[1]{%
	\begin{oldthebibliography}{#1}%
		\setlength{\itemsep}{-0.5ex}%
	}%
	{%
	\end{oldthebibliography}%
}

\end{document}